\documentclass[notitlepage,pallis,twoside]{wpallis}
\usepackage{rotating}
\usepackage{epsfig}
\usepackage{latexsym}
\usepackage{euscript}
\usepackage{amssymb}
\usepackage{times}
\usepackage{slashed,cite}
\usepackage{upgreek}
\usepackage{amsmath}
\usepackage{amsfonts}
\usepackage{amsbsy}
\usepackage{amscd}
\usepackage{bbm}
\usepackage{fancyh}
\usepackage[l]{floatflt}
\numberwithin{equation}{section}

\newcommand{\eec}{\end{center}}
\newcommand{\bec}{\begin{center}}

\newcommand{\eem}{\end{matrix}}
\newcommand{\bem}{\begin{matrix}}
\newcommand{\eeq}{\end{equation}}
\newcommand{\beq}{\begin{equation}}
\newcommand{\ba}{\begin{array}}
\newcommand{\ea}{\end{array}}
\newcommand{\bea}{\begin{eqnarray}}
\newcommand{\eea}{\end{eqnarray}}
\newcommand{\baq}{\begin{eqnarray}}
\newcommand{\eaq}{\end{eqnarray}}
\newcommand{\beqs}{\begin{subequations}}
\newcommand{\eeqs}{\end{subequations}}
\newcommand{\bel}{\begin{align}}
\newcommand{\eeel}{\begin{align}}

\newcommand\eqs[2]{Eqs.~(\ref{#1}) and (\ref{#2})}

\newcommand\eqss[3]{Eqs.~(\ref{#1}), (\ref{#2}) and (\ref{#3})}

\newcommand{\ftn}{\footnotesize}

\newcommand{\ssz}{\scriptsize}

\newcommand{\GeV}{{\mbox{\rm GeV}}}

\newcommand{\sFref}[2]{Fig.~\ref{#1}-{\ftn\sf ({#2})}}
\newcommand{\sEref}[2]{Eq.~(\ref{#1}{\ftn\sf {#2}})}

\def\llgm{\left\lgroup}
\def\rrgm{\right\rgroup}
\def\lf{\left(}
\def\rg{\right)}

\newcommand{\Hhi}{\ensuremath{H_{\rm HI}}}
\newcommand{\Vhi}{\ensuremath{V_{\rm HI}}}
\newcommand{\Vhio}{\ensuremath{V_{\rm HI0}}}
\newcommand{\dV}{\ensuremath{\Delta V_{\rm HI}}}

\newcommand{\Dex}{\ensuremath{\Delta_{1*}}}

\newcommand{\Ve}{\ensuremath{V}}

\newcommand{\Ggut}{\ensuremath{G_{\rm GUT}}}
\newcommand{\Gba}{\ensuremath{G_{21}}}

\newcommand{\Glr}{\ensuremath{G_{\rm LR}}}
\newcommand{\what}{\ensuremath{\widehat}}

\newcommand{\ks}{\ensuremath{k_\star}}
\newcommand{\ns}{\ensuremath{n_{\rm s}}}
\newcommand{\As}{\ensuremath{A_{\rm s}}}
\newcommand{\Ns}{\ensuremath{N_{\star}}}
\newcommand{\as}{\ensuremath{\alpha_{\rm s}}}
\newcommand{\Trh}{\ensuremath{T_{\rm rh}}}

\newcommand{\sgc}{\ensuremath{\sigma_{\rm c}}}

\newcommand{\sg}{\ensuremath{\sigma}}
\newcommand{\sgx}{\ensuremath{\sigma_\star}}
\newcommand{\sgf}{\ensuremath{\sigma_{\rm f}}}

\newcommand{\fp}{\ensuremath{f_p}}
\newcommand{\fpx}{\ensuremath{f_{p\star}}}
\newcommand{\fpc}{\ensuremath{f_{p\rm c}}}

\newcommand{\phc}{\ensuremath{\Phi}}
\newcommand{\phcb}{\ensuremath{\Phi^*}}

\newcommand{\pha}{\ensuremath{\phi_1}}
\newcommand{\phb}{\ensuremath{\phi_2}}
\newcommand{\vpha}{\ensuremath{\varphi_1}}
\newcommand{\vphb}{\ensuremath{\varphi_2}}
\newcommand{\vphii}{\ensuremath{\varphi_\alpha}}
\newcommand{\phii}{\ensuremath{\phi_\alpha}}

\newcommand{\ld}{\ensuremath{\lambda}}

\newcommand{\kp}{\ensuremath{\kappa}}
\newcommand{\se}{\ensuremath{\widehat\sigma}}

\newcommand{\sex}{\ensuremath{\widehat{\sigma}_*}}
\newcommand{\sef}{\ensuremath{\widehat{\sigma}_{\rm f}}}
\renewcommand{\sec}{\ensuremath{\widehat{\sigma}_{\rm c}}}

\newcommand{\eph}{\ensuremath{\epsilon}}
\newcommand{\ith}{\ensuremath{\eta}}

\newcommand\vev[1]{\langle {#1} \rangle}
\newcommand\vevi[1]{\langle {#1} \rangle_{\rm HI}}
\newcommand\mtta[2]{\mbox{
$\llgm\bem #1 \cr #2 \eem\rrgm$}}

\newcommand{\dm}{\ensuremath{\widetilde m}}
\renewcommand{\th}{\ensuremath{\theta}}
\newcommand{\Nr}{\ensuremath{{\sf N}_{\rm G}}}
\newcommand{\mP}{\ensuremath{m_{\rm P}}}
\newcommand{\suba}{\ensuremath{SU(2)\times U(1)}}

\newcommand{\msn}{\ensuremath{\what m_{\sg}}}
\newcommand{\mdp}{\ensuremath{m_{\delta\pha}}}
\newcommand{\mth}{\ensuremath{\what m_{\th}}}
\newcommand{\arctanh}{\ensuremath{\rm arctanh}}

\def\Ka{K\"{a}hler potential}

\def\sub{subplanckian}

\newcommand{\plk}{{\it Planck}}
\def\sub{sub-Planckian~}
\newcommand{\thi}{{THI}}
\newcommand{\ehi}{{EHI}}
\newcommand{\ethi}{{EHI and THI}}

\def\bbet{{\bar\beta}}
\def\al{{\alpha}}
\def\bt{{\beta}}

\begin{document}

\thispagestyle{empty}
%%%%%%%%%%%%%%%

\title[]{\Large\bfseries\scshape E- \& T-Model Hybrid Inflation}

\author{\large\bfseries\scshape C. Pallis}
\address[] {\sl  Laboratory of Physics, Faculty of
Engineering, \\ Aristotle University of Thessaloniki, \\
GR-541 24 Thessaloniki, GREECE\\ {\tt kpallis@auth.gr}}

\begin{abstract}{\small {\bfseries\scshape Abstract} \\
\par We consider the impact of a kinetic pole of order one or two
on the non-supersymmetric model of hybrid inflation. These poles
arise due to logarithmic \Ka s which control the kinetic mixing of
the inflaton field and parameterize hyperbolic manifolds with
scalar curvature related to the coefficient $(-N)<0$ of the
logarithm. Inflation is associated with the breaking of a local
\suba\ symmetry, which does not produce any cosmological defects
after it, and remains largely immune from the minimal possible
radiative corrections to the inflationary potential. For $N=1$ and
equal values of the relevant coupling constants, $\lambda$ and
$\kappa$, the achievement of the observationally central value for
the scalar spectral index, $\ns$, requires the mass parameter,
$m$, and the symmetry breaking scale, $M$, to be of the order of
$10^{12}~\GeV$ and $10^{17}~\GeV$ respectively. Increasing $N$
above unity the tensor-to-scalar ratio $r$ increases above $0.002$
and reaches its maximal allowed value for $N\simeq 10-20$. }
%
%\\ \\{\ftn \sc Keywords}: {\ftn Cosmology};
%{\ftn\sc PACS codes:} {\ftn 98.80.Cq}
\end{abstract} \maketitle
%\publishedin{{\sl  Phys. Lett. B} {\bf 692}, 287 (2010)}

\thispagestyle{empty}

\setcounter{page}{1} \pagestyle{fancyplain}

%\addtolength{\headheight}{.5cm}

\rhead[\fancyplain{}{ \bf \thepage}]{\fancyplain{}{\sl\small E- \&
T-Model Hybrid Inflation}} \lhead[\fancyplain{}{\sl
\leftmark}]{\fancyplain{}{\bf \thepage}} \cfoot{}

\vspace*{-.cm}

\section{Introduction}\label{intro}

It is widely believed that the introduction of
\emph{supersymmetry} ({\sf\ftn SUSY}) and its local extension --
\emph{supergravity} ({\sf\ftn SUGRA}) -- can alleviate the
shortcomings of \emph{Standard Model} ({\sf\ftn SM})  and provide
a safe framework for building a variety of inflationary models --
see e.g. \cref{sugrareview}. However, we have to accept that there
is no direct experimental confirmation of SUSY until now
\cite{lhc}. On the contrary, there is a strong observational
evidence in favor of the inflationary paradigm \cite{plcp,plin}.
Consequently, it is worthwhile to build inflationary models
compatible with the observations, even without the presence of
SUSY.

One of those models -- for reviews see \cref{review} -- is
undoubtedly \emph{Hybrid Inflation} ({\sf\ftn  HI}) \cite{hybrid}
which elegantly combines a period of inflation with a phase
transition at its termination. It is termed ``hybrid'' because the
inflationary vacuum energy density is provided by a waterfall
field, different from the slowly rolling inflaton field. The model
allows for values of the relevant coupling constants larger than
those used in chaotic inflation, and can be successfully handled
with subplanckian values for the inflaton field. It can be, also,
embedded in several \emph{Grand Unified Theory} ({\sf\ftn GUT})
schemes  -- mainly in the SUSY framework \cite{fhi,su5,notd} --
and nicely connects inflationary cosmology with particle physics.
A very intriguing possibility of these embeddings is the
production of topological defects at its end via the Kibble
mechanism \cite{kibble}. Between them, the metastable cosmic
strings are currently of special interest since they may decay
generating a stochastic background of gravitational waves
\cite{kainano} and interpreting, thereby, recent results
\cite{nano} -- for other sources of gravitational waves'
production after the end of HI see \cref{spanos, dim}.

On the other hand, HI suffers from the problem of the enhanced
(scalar) spectral index $\ns$ which turns out to be, mostly, well
above the present data \cite{pl20, gws1, gws} -- for another point
of view see \cref{hi22}. Indeed, $\ns$ within tree level HI
exceeds unity \cite{hybrid} and only in a minor and tuned region
of parameters \cite{nmi} a reconciliation with data can be
achieved. Inclusion of \emph{radiative corrections} {(\sf\ftn
RCs)} \cite{coleman} due to a possible coupling of the inflaton to
fermions \cite{hirc} or a conveniently selected non-minimal
coupling to gravity \cite{nmi,nmi10} can reconcile the model's
predictions with observations -- for another mechanism possibly
applicable in non-SUSY HI see \cref{suphi}. However, the situation
remains problematic since in both aforementioned cases, HI of
hilltop type \cite{lofti} is obtained and so an unavoidable tuning
of the initial conditions emerges. Nonetheless, it is by now
well-known \cite{terada,so,epole,etpole} that the presence of a
pole in the kinetic term of the inflaton facilitates the
achievement of observation-friendly chaotic inflation. It would
be, therefore, interesting to investigate if this technique can be
also applied for HI improving, thereby, its compatibility with the
observations -- for a similar recent work see \cref{hlinde}.

We find out that two types of HI can be formulated depending on
the order $p$ of the inflaton kinetic pole \cite{linde21,
ellis21}. For $p=1$ we obtain \emph{E-model HI} ({\sf\ftn EHI})
whereas for $p=2$, \emph{T-model HI} ({\sf\ftn THI}) arises. The
terms are coined in analogy to \emph{E-model} (chaotic)
\emph{inflation} ({\sf\ftn EMI}) \cite{alinde} (or
$\alpha$-Starobinsky model \cite{eno7}) and \emph{T-model
inflation} ({\sf\ftn TMI}) \cite{tmodel}. As in the latter cases
\cite{etpole}, EHI and THI can be relied on \Ka s and so the
relevant particle models can be established as non-linear sigma
models with specific geometry of the moduli space. In both cases,
the minimal possible RCs \cite{coleman} to the inflationary
potential are considered and subplanckian values for the initial
(non-canonically normalized) inflaton field are required.
Unfortunately, our scheme is inconsistent with the formation of
cosmic strings after HI, since the relevant scale $M$ of the phase
transition turns out to be a little lower than the reduced Planck
scale, $\mP= 2.44\cdot 10^{18}~\GeV$, and so the tension of the
cosmic strings \cite{kainano} would be unacceptably large.
Therefore, the version of HI with the waterfall field charged
under the continuous group $U(1)$ with the lowest possible
dimensionality is not a representative working example for our
setup. However, if there is a $U(1)$ factor in the initial and the
remaining gauge group, with the waterfall field arranged in a
convenient representation, then the production of cosmic strings
(and monopoles) after HI can be eluded \cite{td}. In our analysis
we adopt the simplest possible scenario for reference. Other ways
to overcome the obstacle of topological defects within HI are
proposed in \cref{hlinde, notd}.

%\tableofcontents\vskip-1.3cm\noindent\rule\textwidth{.4pt}\\

Below, in \Sref{hi}, we describe how we can formulate these
versions of HI. The dynamics of the resulting inflationary models
is studied in \Sref{inf} and these are tested against observations
in \Sref{res}. Finally, \Sref{con} summarizes our conclusions. In
Appendix~A we explore the possibility for generating different
moduli kinetic mixings allowed by generic \Ka s. Throughout the
text, the subscript $,\chi$ denotes derivation \emph{with respect
to} ({\sf\small w.r.t}) the field $\chi$, charge conjugation is
denoted by a star ($^*$) and we use units where $\mP = 1$ unless
otherwise stated.

\section{Models' Setup}\label{hi}

Since E and T models are introduced by means of a non-minimal
kinetic mixing \cite{etpole}, we find it convenient to establish
their combination with HI taking as reference the non-linear sigma
models. In these the kinetic mixing is controlled by a metric
$K_{\al\bbet}$ defined on the moduli space. Here we employ two
(complex) scalar fields $Z^\al$ with $\al=1,2$ -- the inflaton
$Z^1=S$ and the waterfall field $Z^2=\Phi$. Therefore, the
relevant lagrangian terms are written as
\beq\label{Saction1} {\cal  L} = \sqrt{-\mathfrak{g}} \lf
K_{\al\bbet}\lf D_\mu Z^\al\rg^\dagger D^\mu Z^{\bbet}
-V(Z^\al)\rg, \eeq
where $\mathfrak{g}$ is the determinant of the background
Friedmann-Robertson-Walker metric $g^{\mu\nu}$ with signature
$(+,-,-,-)$. We further assume that $K_{\al\bbet}$ originates from
a K\"ahler potential $K$ -- as in the context of SUGRA --
according to the definition
\beq \label{kdef} K_{\al\bbet}={K_{,Z^\al
Z^{*\bbet}}}>0\>\>\>\mbox{with}\>\>\>K^{\bbet\al}K_{\al\bar
\gamma}=\delta^\bbet_{\bar\gamma}.\eeq
We below, in \Sref{hi1} and \ref{hi2}, we respectively specify the
kinetic terms and the inflationary potential of our models.

\subsection{Kinetic Mixing}\label{hi1}

The adopted $K$'s include two contributions, one for the inflaton,
$K_{\rm I}$, and one for the waterfall field, $K_{\rm W}$. Namely,
we set
\beq \label{Ket}  K=K_{\rm I}+K_{\rm W}\>\>\mbox{with}\>\>K_{\rm
W}=|\Phi|^2\>\>\mbox{and}\>\>K_{\rm I}=
\begin{cases}
-2N\ln(1-(S+S^*)/2) &\mbox{for EHI},\\
-(N/2)\ln(1-|S|^2)  &\mbox{for THI}\,.
\end{cases}\eeq
We see that no mixing between the inflaton and waterfall sectors
exists and the non-zero elements of the relevant metric are
calculated to be
\beq \label{Ksset}
 K_{\phc\phcb}=1\>\>\mbox{and}\>\>K_{SS^*}=\frac{N}{2}\cdot
\begin{cases}
(1-(S+S^*)/2)^{-2} &\mbox{for EHI},\\
(1-|S|^2)^{-2}  &\mbox{for THI}\,.
\end{cases}\eeq
%\eeq
%
For both models, $K_{\rm W}$ parameterizes flat manifold --
contrary to the setting in \cref{hlinde, dim} -- whereas the
geometry induced by $K_{\rm I}$ is hyperbolic, i.e., the curvature
of the moduli space is negative. In particular, the scalar
curvatures associated with $K_{\rm W}$ and $K_{\rm I}$
respectively are
\beq \label{Rs}{\cal R}_{\rm W}=0 \>\>\mbox{and}\>\>{\cal R}_{\rm
I}=-K^{SS^*}\partial_S\partial_{S^*}\ln
K_{SS^*}=-({1}/{N})\begin{cases}
1&\mbox{for EHI},\\
4&\mbox{for THI}.
\end{cases}\eeq
More specifically, $K_{\rm I}$ parameterizes the coset space
$SU(1,1)/U(1)$ in the case of THI, whereas for EHI the space
generated by $K_{\rm I}$ is invariant under the set of
transformations related to the group $U(1,1)$ -- see \cref{epole}.

In \Eref{Saction1} we include the covariant derivatives for the
scalar fields $D_\mu Z^\al$ which are given by
\beq \label{dz} D_\mu Z^\al=\partial_\mu Z^\al+ig A^{\rm a}_\mu
T^{\rm a}_{\al\bt} Z^\bt,\eeq
with $A^{\rm a}_\mu$ being the vector gauge fields, $g$ a gauge
coupling constant and $T^{\rm a}$ with ${\rm
a}=1,...,\mbox{\sf\ftn dim}\Ggut$ the generators of the gauge
group $\Ggut$ with dimensionality $\mbox{\sf\ftn dim}\Ggut$. To
avoid the production of cosmic defects after the end of HI we have
to select $\Ggut$ according to the guidelines of \cref{td}. For
definiteness, we here assume that $\phc$ belongs to the $({\bf 2},
1)$ representation of the group $\Gba=SU(2) \times U(1)$.
Interestingly enough, this group may be identified with the
$SU(2)_{\rm R} \times U(1)_{B-L}$ part of a realistic (minimal)
version \cite{mlr} of the GUT based on the group $\Glr=SU(3)_{\rm
C}\times SU(2)_{\rm L} \times SU(2)_{\rm R} \times U(1)_{B-L}$.

If we use the following parameterizations for the two fields
\beq\label{para} S=\sigma
e^{i\th}\>\>\mbox{and}\>\>\Phi=\frac{1}{\sqrt{2}}\mtta{\pha+i\vpha}{\phb+i\vphb},\eeq
we can introduce he canonically normalized (hatted) fields during
HI as follows
\beq \label{Kc} \vevi{K_{\al\bbet}}\dot Z^\al \dot Z^{*\bbet}
\simeq \frac12\lf\dot{\widehat \sg}^2+\dot{\widehat
\th}^2+\dot{\phii}^2+\dot{\vphii}^2\rg\>\>\mbox{for}\>\>\al=1,2
\eeq
where $K_{\al\bbet}$ is given by \Eref{Ksset} and assures the
canonical normalization of $\phii$ and $\vphii$. The symbol
``$\vevi{Q}$" denotes the value of a quantity $Q$ during HI, where
all the fields besides $\sg$ are set equal to zero -- see below --
and dot stands for derivation w.r.t the cosmic time. From
\Eref{Kc} the remaining fields $\se$ and $\what{\theta}$ are
identified as
\beq \label{je}
\frac{d\se}{d\sg}=J=\frac{\sqrt{N}}{\fp}~~\Rightarrow~~\se=\begin{cases}
-\sqrt{N}\ln(1-\sg) &\mbox{for}~~p=1,\\
\sqrt{N}\arctanh{\sg} &\mbox{for}~~p=2
\end{cases}\>\>\mbox{and}\>\>
\widehat{\theta}\simeq J\sg\theta, \eeq
where $\fp=1-\sg^p$ with $p=1,2$. Note that $\se$ gets increased
above unity for $\sgc\leq\sg<1$, facilitating, thereby, the
attainment of HI with \sub\ $\sg$ values.

\subsection{Inflationary Potential }\label{hi2}

In \Eref{Saction1} the inflationary potential for our models is
also incorporated. Its form is
\beq V(S,\phc) = \kappa^2\left(|\Phi|^2-M^2\right)^2+\frac12
m^2|S|^2+\ld^2|\Phi|^2|S|^2+\frac18\dm|S-S^*|^2,\label{Vhi1}\eeq
where $M$, $m$ and $\dm$ are mass parameters whereas $\kappa$ and
$\lambda$ are dimensionless coupling constants. The last unusual
term is adopted to provide the angular mode of $S$, $\th$, with
mass, as we see below. The global minima of $V$ lie at the
direction
\beq
\label{vev}\vev{\pha}=\pm\sqrt{2}M\>\>\mbox{and}\>\>\vev{\sg}=\vev{\phb}=\vev{\vphii}=\vev{\th}=0\eeq
and ensure a spontaneous breaking of the local symmetry $\Gba$
down to a $U(1)'$. The scalar spectrum of the theory is composed
by three real scalars $\sg$, $\delta \pha=\pha-\vev{\pha}$ and
$\th$ with masses correspondingly
\beq \label{mass} \msn=\sqrt{(m^2+2\ld^2 M^2)/N},\>\>\mdp=2\kp
M\>\>\mbox{and}\>\>\mth=\dm/\sqrt{N}.\eeq
Thanks to the selected representation of $\phc$ in \Eref{para}
neither cosmic strings nor monopoles are left behind the breakdown
of \Gba \cite{td}. The truncated form of $V$ with all the fields,
but $\sg$ and $\pha$, set at the origin reduces to its well-known
form \cite{hybrid} within HI, i.e.,
\begin{equation}\label{Vhi2}
V \left(\pha,\sigma\right) = \kappa^2 \left(\frac{\pha^2}{2}-M^2
\right)^2+\frac{m^2}{2}\sigma^2+\frac{\lambda^2}{2}\pha^2
\sigma^2\,.
\end{equation}

Besides the phase transition above, $V$ in \Eref{Vhi1} gives rise
to HI. This is, because it possesses an almost $\sg$-flat
direction at
\beq
\label{vevi}\vevi{\phii}=\vevi{\vphii}=\vevi{\th}=0\>\>\mbox{with}\>\>\sg\geq\sgc
= \sqrt{2}\kappa M/\lambda\,.\eeq
The last inequality here assures the stability of the inflationary
path w.r.t the fluctuations of $\phii$ and $\vphii$ during HI.
Indeed, as shown in \Tref{tab1} -- where the mass squared spectrum
of the model along the trough in \Eref{vevi} is arranged -- the
positivity of $m^2_{\Phi}$ is protected by the last inequality in
\Eref{vevi}. From \Tref{tab1} we can also verify that $\th$
acquires mass thanks to the last term of \Eref{Vhi1} and confirm
that $\what m^2_{\th}\gg\Hhi^2$ and $\what m^2_{\phc}\gg\Hhi^2$
for $\sgc<\sg\leq\sgx$. Here we define
\beq \mbox{\ftn\sf
(a)}\>\>\Hhi=\sqrt{\Vhio/3}\>\>\>\mbox{with}\>\>\>\mbox{\ftn\sf
(b)}\>\>\Vhio=\kappa^2M^4\simeq V(S,0)\label{Hhi}\eeq
the almost constant potential density during HI. Note that no
massive gauge bosons exist in the particle spectrum since $\Gba$
is unbroken during HI. Using the derived spectrum we can compute
the one-loop RCs (to the tree-level potential) $\dV$  which can be
written as \cite{coleman}
\beq \label{Vrc} \dV ={1\over64\pi^2}\lf \widehat
m_{\th}^4\ln{\widehat m_{\th}^2\over\Lambda^2} +2\Nr \widehat
m_{\phi}^4\ln{\widehat m_{\phi}^2\over\Lambda^2} \rg.\eeq
Here, $\Nr=2$ is the dimensionality of the representation of
$\phc$ and $\Lambda$ is a renormalization-mass scale which, is
determined by requiring $\dV(\sgx)=0$ or $\dV(\sgc)=0$. This
determination of $\Lambda$ renders our results practically
independent of $\Lambda$ in the major parameter space of the
models. Indeed, these can be derived exclusively by using the tree
level $\Vhi=V(S,0)$ with the various quantities evaluated at
$\Lambda$ -- the relevant renormalization-group running is
expected to be negligible  -- see \Sref{res}. On the other hand,
as shown in the same section, $\dV$ is capable to ruin the
successful inflationary predictions for $\kp\gtrsim0.003$. This
effect may be avoided if extra contributions with opposite sign
are included into $\dV$ due to possible coupling of $S$ or $\phc$
to fermions (such as right-handed neutrinos) -- see \cref{hirc}.

In conclusion, the inflationary potential for \ethi\ is
\beq \Vhi=\Vhio+{1\over2}m^2\sg^2+\dV\label{Vhi}\eeq
with the canonical normalization of $\sg$ in \Eref{je} being taken
into account and differentiating the models between each other.

\renewcommand{\arraystretch}{1.2}
\begin{table}[!t]
\bec\begin{tabular}{|c||c|c|c|}\hline {\sc Fields}&{\sc Eigen-}&
\multicolumn{2}{c|}{\sc Masses }\\ &{\sc states}& \multicolumn{2}{c|}{\sc Squared}\\
\hline\hline
$1$ real scalar &$\widehat \theta$ & $\widehat m^2_{\theta}$& \multicolumn{1}{|c|}{$\dm^2\fp^2/N$}\\
$2\Nr$ real scalars &$\phii,~\vphii$ & $\what m^2_{
\Phi}$&\multicolumn{1}{|c|}{$\ld^2\sg^2-2\kp^2M^2$}\\\hline
\end{tabular}\eec
\hfill \vchcaption[]{\sl Mass spectrum for EHI $(p=1)$ and THI
($p=2$) along the inflationary path in \Eref{vevi}.}\label{tab1}
\end{table}\renewcommand{\arraystretch}{1}

\section{Inflation Analysis}\label{inf}

A period of slow-roll EHI or THI is controlled by the strength of
the slow-roll parameters which can be derived by applying the
standard formulae -- see e.g. \cref{review}. Note that the
derivation can be performed employing $\Vhi$ in \Eref{Vhi} and $J$
in \Eref{je}, without express explicitly $\Vhi$ in terms of $\se$
-- see e.g. \cref{nmi}. Namely, we find
\beqs\beq\label{sr}\epsilon= {1\over2}\left(\frac{V_{{\rm
HI},\se}}{\Vhi}\right)^2\simeq \frac{m^4\fp^2\sg^2}{2N\Vhio^2}
\>\>\>\mbox{and}\>\>\>\eta = \frac{V_{{\rm HI},\se\se}}{\Vhi}
\simeq m^2\fp\frac{\fp-p\sg^p}{N\Vhio} \cdot\eeq
In the expressions above we replace $\Vhi$ from \sEref{Hhi}{b}
whereas $V_{{\rm HI},\se}$ and $V_{{\rm HI},\se\se}$ are obtained
by performing derivations of \Eref{Vhi}. As $m$ increases beyond
$5\cdot10^{-6}$ the approximation above becomes less accurate. It
can be verified numerically that
\beq\label{sr1}\eph(\sgc)\leq1\>\>\>\mbox{and}\>\>\>
\ith(\sgc)\leq1\eeq\eeqs
and so HI is terminated prematurely -- i.e., for $\sgf=\sgc$ -- in
the major parametric space of our model.

To protect our scenario from the production of extra e-foldings
\cite{hybrid, hlinde, bjorn} during the waterfall regime, we take
into account the behavior of $\sg$ and $\pha$ after the time
$\Hhi^{-1}$ from the moment when $\sg=\sgc$, following the
approach in \cref{hybrid,hlinde}. The absolute value of the $\pha$
effective mass after this time can be estimated as
\beq\label{Dm} \Delta
m_{\pha}^2\simeq2\ld^2\Delta\sg\sgc\>\>\mbox{where}\>\>\Delta\sg=3\sqrt{2}{m^2\mP^2}/{\ld\kp
M^3}\eeq
is the reduction of the $\sg$ value below $\sgc$ as can be
computed by the inflationary equation of motion. No sizable amount
of inflation occurs during the waterfall regime, if
\beq \label{kbou} {\Delta
m_{\pha}^2}/{\Hhi^2}\simeq36{m^2\mP^4}/{\kp^2
M^6}>1\>\>\Rightarrow\>\>\kp\lesssim{6m\mP^2}/{M^3},\eeq
where $\Hhi^2$ is given by \sEref{Hhi}{a}. This condition is
roughly identical with that obtained in more accurate computations
\cite{bjorn} and can be easily met in our scenario -- see
\Sref{res}.

The number of e-foldings $\Ns$ that the pivot scale $\ks=0.05/{\rm
Mpc}$ experiences during EHI or THI can be calculated through the
relation
\begin{equation}
\label{Nhi} \Ns=\int_{\sef}^{\sex} d\se\frac{\Vhi}{V_{\rm
HI,\se}}=\frac{N\Vhio}{pm^2}\lf\ln\frac{\fpc}{\fpx}+p\ln\frac{\sgx}{\sgc}+\frac1\fpx-\frac1{\fpc}\rg,\eeq
where $\fpx=\fp(\sgx)$ and $\fpc=\fp(\sgc)$ with $\sgx~[\sex]$
being the value of $\sg~[\se]$ when $\ks$ crosses the inflationary
horizon. Taking into account that $1\simeq\sgx\gg\sgf=\sgc$, we
can find the following approximate version of the exact formula
above which can be analytically solved w.r.t $\sgx$ as follows
\begin{equation} \label{sgx}
\Ns\simeq\frac{N\Vhio}{pm^2}\lf\frac{1}{\fpx}-1-p\ln\sgc\rg\>\>\Rightarrow\>\>\sgx\simeq
\lf1-\frac{N\Vhio}{N\Vhio(1+p\ln\sgc)+\Ns pm^2}\rg^{1/p}\,.
\end{equation}
The elimination of the terms $p\ln\sgx-\ln\fpx$ can be justified
by the fact that their contribution to $\Ns$ in \Eref{Nhi} is
subdominant for the $\sg$ values in the range $0.5\leq\sgx<1$
encountered in our numerics.

The amplitude $\As$ of the power spectrum of the curvature
perturbations generated by $\sg$ at the scale $\ks$ is estimated
as follows
\beq \label{Prob}\sqrt{\As}=\: \frac{1}{2\sqrt{3}\, \pi} \;
\frac{\Ve_{\rm HI}(\sex)^{3/2}}{|\Ve_{\rm
HI,\se}(\sex)|}\simeq\frac{\sqrt{N}\Vhio^{3/2}}{2\sqrt{3}\pi\sgx
m^2\fpx}\,\cdot  \eeq
Inserting \Eref{sgx} into the previous equation and expanding the
resulting relation in powers of $\Ns$ we can obtain a rough
estimation of $\kp$ as follows
\beq \label{lda}\sqrt{\As}
\simeq\:\frac{\sqrt{N}\Vhio^{3/2}}{2\sqrt{3}\pi
m^2}\lf1+\frac{N\Vhio}{p^2m^2\Ns}\rg\>\>\Rightarrow\>\>
\kp\simeq\lf\frac{2\pi\sqrt{3\As}m^2}{M^6N^{1/2}}\rg^{1/3}\cdot\eeq

We, also, calculate the remaining inflationary observables via the
relations
\beqs\baq \label{ns} && \ns=\: 1-6\epsilon_\star\ +\
2\eta_\star\simeq
1-\frac{2}{\Ns}-\frac{3N}{p^2\Ns^2}+\frac{\Vhio}{m^2\Ns^2}\lf{2N\over p^2}+{4N\over p}+N\ln\sgc^2\rg\,,\\
&& \label{as} \as =\:{2\over3}\left(4\eta_\star^2-(n_{\rm
s}-1)^2\right)-2\xi_\star\simeq-\frac{2}{\Ns^2}-\frac{6N}{p^2\Ns^3}+\frac{N\Vhio}{m^2\Ns^3}\lf{4\over
p^2}+{10\over p}+\ln\sgc^4\rg,\>\>\>\>\>\>\>\>\\
&& \label{rs} r=16\epsilon_\star\simeq
\frac{8N}{p^2\Ns^2}-\frac{16N^2\Vhio}{p^2m^2\Ns^3}\lf{1\over
p^2}+{1\over p}+\ln\sgc\rg,\eaq\eeqs
where $\xi={\Ve_{\rm HI,\widehat\sg} \Ve_{\rm
HI,\widehat\sg\widehat\sg\widehat\sg}/\Ve_{\rm HI}^2}$, the
variables with subscript $\star$ are evaluated at $\sg=\sgx$ and
the approximate expressions are obtained by expanding the exact
result in powers of $1/\Ns$. A clear dependence of the observables
on the parameters of $\Vhi$ in \Eref{Vhi} arises which is expected
to modify considerably the -- dominating at leading order in the
expansions above -- predictions of the pure E- and T-model
inflation \cite{linde21, ellis21}.

We should, finally, note that the results above can be deliberated
from our ignorance about the Planck-scale physics, if we impose
two additional theoretical constraints on our models:
\beq \label{subP}\mbox{\ftn\sf (a)}\>\>\Vhi(\sgx)^{1/4}\leq1
\>\>\>\mbox{and}\>\>\>\mbox{\ftn\sf (b)}\>\>\sgx\leq1.\eeq
As we show below, the first from the inequalities above is easily
satisfied in our set-up for $\kp$ and $M$ values consistent with
\Eref{lda} -- see also \Sref{res}. \sEref{subP}{b} is fulfilled by
construction since the $K$'s in \Eref{Ket} with the
parameterizations in \Eref{para} induce a kinetic pole for $\sg=1$
and so HI takes place for $\sg<1$. The introduction of a possible
parameter multiplying $S$ in \Eref{Ket} can be absorbed by a
reparametrization of the free parameters of the model -- see e.g.
\cref{so}.

\section{Numerical Results}\label{res}

The outputs of the analysis above can be refined numerically and
employed in order to delineate the available parameter space of
the models. In particular, we confront the quantities in
\eqs{Nhi}{Prob} with the observational requirements \cite{plin}
\beq\mbox{\ftn\sf
(a)}\>\>\Ns\simeq61.3+\frac{1}{12}\ln\frac{\pi^2g_{\rm
rh*}\Trh^4}{30\Vhi(\sgf)}\>\>\>\mbox{and}\>\>\>\mbox{\ftn\sf
(b)}\>\>\sqrt{\As}\simeq4.588\cdot10^{-5}.\label{ntot}\eeq
%}{
In deriving \sEref{ntot}{a} we assume that HI is followed in turn
by a oscillatory phase, with mean equation-of-state parameter
$w_{\rm rh}\simeq0$, radiation and matter domination. We expect
that $w_{\rm rh}=0$, corresponding to quadratic potential,
approximates rather well $\Vhi$ for $\sg\ll\sgc$. Motivated by
implementations  of non-thermal leptogenesis \cite{inlept}, which
may follow HI, we set $\Trh\simeq4.1\cdot10^{-10}$ for the reheat
temperature. Also $g_{\rm rh*}=106.75$ is the energy-density
effective number of degrees of freedom which corresponds to SM
spectrum respectively.

As regards the remaining observables, we take into account the
latest data from \emph{Planck} (release 4) \cite{pl20}, baryon
acoustic oscillations, \emph{Cosmic Microwave Background} {\ftn
\sf (CMB)} lensing and BICEP/{\it Keck} \cite{gws1}. Adopting the
most updated fitting in \cref{gws} we obtain approximately the
following allowed margins
\begin{equation}  \label{data}
\mbox{\ftn\sf
(a)}\>\>\ns=0.965\pm0.0074\>\>\>\mbox{and}\>\>\>\mbox{\ftn\sf
(b)}\>\>r\leq0.032,
\end{equation}
at 95$\%$ \emph{confidence level} ({\sf\ftn c.l.}) with
$|\as|\ll0.01$. The constraint on $|\as|$ is readily satisfied
within the whole parameter space of our models. Enforcing
\sEref{ntot}{a} and {\sf\ftn (b)} we can restrict $\kp$ and $\sgx$
and compute the models' predictions via \eqss{ns}{as}{rs}, for any
selected $N$, $\ld/\kp$, $m$ and $M$ -- see \Eref{Vhi1}. Note that
$\dm$ is totally irrelevant for the computation and can be fixed
at a value (e.g., $\dm=10m$) throughout so that $\what
m_\th\gg\Hhi$. With this arrangement, $\dV$ in \Eref{Vrc} is
totally dominated by its second term in the equation above. Given
that $\Lambda$ in \Eref{Vrc} can be determined self-consistently
as mentioned in \Sref{hi2}, $\dV$ is calculated using as inputs
the free parameters of the tree-level potential in \Eref{Vhi1}.
Compared to the original model of HI \cite{hybrid}, we here
employ one more free parameter, $N$, related to the non-minimal
kinetic terms adopted in \Eref{Ksset}.

We embark on the presentation of our results by comparatively
plotting the variation of $\Vhi$ as a function of $\sg$ in
\sFref{fig1}{a} and $\se$ in \sFref{fig1}{b} for \ehi\ (solid
lines) and \thi\ (dashed lines). In both cases we set $\ld=\kp$
and $M=0.1$ and achieve the central $\ns$ in \sEref{data}{a} by
selecting $m=4.12\cdot10^{-6}$ [$m=2.04\cdot10^{-6}$] for \ehi\
[\thi]. The corresponding values of $\kp$, $\Ns$, $\sgx$  and $r$
are listed in the third, fourth, fifth and sixth leftmost column
of the Table below the graphs. These results are obtained by our
numerical code taking into account exact expressions for $J, \Vhi$
and the other observables -- i.e., Eqs.~(\ref{je}), (\ref{Vhi}),
(\ref{Nhi}), and the leftmost equalities in Eqs.~(\ref{ns}),
(\ref{as}) and (\ref{rs}). These values can be approximated by
those obtained by employing the formulas of \Sref{inf} -- i.e.,
Eqs.~(\ref{sgx}), (\ref{lda}), and the rightmost equalities in
Eqs.~(\ref{ns}), (\ref{as}) and (\ref{rs}). The outputs of this
computation are displayed in the five rightmost columns of the
Table in \Fref{fig1}. Note that in the case of analytic
expressions we prefer to compare $\sgx$ derived by \Eref{sgx} with
its numerical value and let $m$ as input parameter. We remark that
$\Ns$ is analytically underestimated in both cases. This fact
leads to an overestimation of $\sgx$ and $\kp$. On the other hand,
$\ns$ and $r$ are closer to their numerical values for \ehi\ than
for \thi. The non-consideration of the second term in \Eref{Vhi}
in the denominators of \Eref{sr} seems to aggravates the error in
analytic expressions of \thi. Despite their low accuracy, though,
the analytical expressions of \Sref{inf} assist us in
understanding the general behavior of the inflationary dynamics.

%In \ehi\ we obtain $\as\simeq-3.5\cdot10^{-4}$ whereas for \thi\
%we get $\as\simeq6\cdot10^{-4}$.

In the plots of \Fref{fig1} we verify the pretty stable mechanism
\cite{tmodel, so, epole} which establishes  EMI and TMI: $\Vhi$
expressed in terms of $\se$ develops a plateau for $\se\gg1$ but
$\sg<1$ -- since $\se$ increases w.r.t $\sg$ as inferred from
\Eref{je}. Indeed, the $\sgx$ values depicted in \sFref{fig1}{a}
-- and arranged in the Table of \Fref{fig1} -- get enhanced
according to \Eref{je}, i.e., $\sex=0.99$ [$1.66$] for \ehi\
[\thi] -- see \sFref{fig2}{b}. Contrary to the standard EMI and
TMI, though, we here observe that {\ftn\sf (i)} the inflationary
path terminates at $\sg=\sgc$ due to the instability in
\Eref{vevi} -- this is the same for both EHI and THI since $M$ and
$\ld/\kp$ are kept fixed in both cases -- and {\ftn\sf (ii)} the
required $\sgx$ does not lie extremely close to the location of
the pole at $\sg=1$ and so the relevant tuning of the initial
conditions, somehow quantified -- cf. \cref{so,epole} -- by the
quantity
\beq \Dex=\left(1 - \sgx\right),\label{dex}\eeq
is less severe here. Namely, we obtain $\Dex=19\%$ [$\Dex=24\%$]
for \ehi\ [\thi], whereas $\Dex\leq10\%$ in all the models of
chaotic inflation studied in \cref{so,epole, etpole}. Also no
proximity is needed between $\sgx$ and $\sgc$ as in the last paper
of \cref{su5}.

%%%%%%%%%%%%%%%%%%%%%%%%%%%%%%%%%%%%%%%%%%%%%%%%%%%%%%%%%%%%%%%%%%%%
\begin{figure}[!t]\vspace*{-.1in}
\hspace*{-.19in}
\begin{minipage}{8in}
\epsfig{file=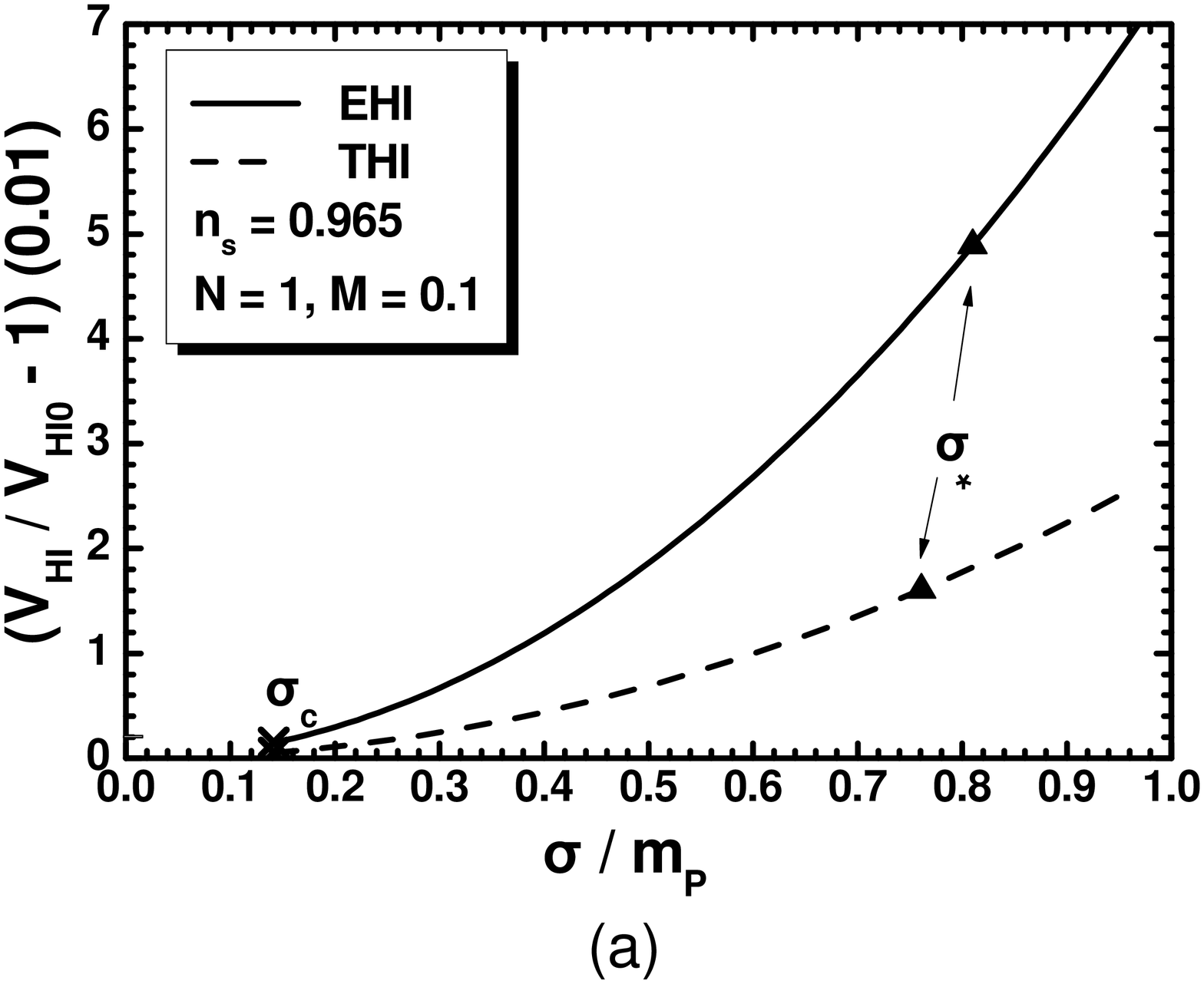,height=3.6in,angle=-90}
\hspace*{-1.2cm}
\epsfig{file=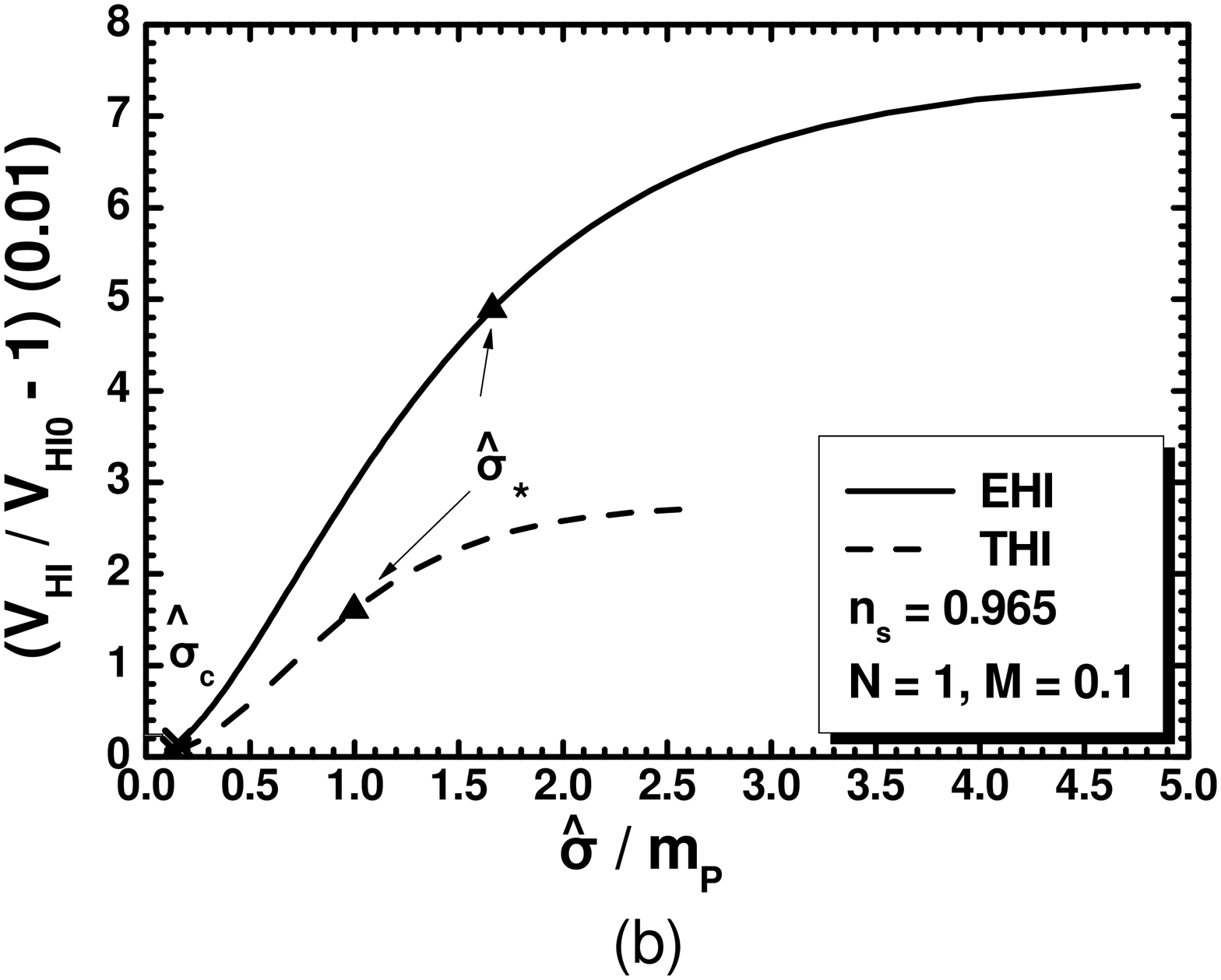,height=3.6in,angle=-90} \hfill
\end{minipage}
\renewcommand{\arraystretch}{1.2}
\bec{\small \begin{tabular}{|c|c||c|c|c|c||c|c|c|c|c|}\hline
{\sc Model}&$m/10^{-6}$&\multicolumn{4}{c||}{\sc Numerical
Values}&\multicolumn{5}{c|}{\sc Analytic Values}\\\cline{3-11}
&&$\kp/0.001$&$\Ns$&$\sgx$&$r/0.001$&$\kp/0.001$&$\Ns$&$\sgx$&$\ns$&$r/0.001$\\\hline\hline
EHI&$4.12$&$1.07$&$50.6$&$0.81$&$3.83$&$2$&$41.7$&$0.85$&$0.965$&$3.1$\\
THI&$2.04$&$0.87$&$50.2$&$0.76$&$2.44$&$1.3$&$47.7$&$0.79$&$0.95$&$1.45$\\\hline
\end{tabular}}\eec
\renewcommand{\arraystretch}{1.0}
\hfill \vchcaption[]{\sl\small Variation of $\Vhi$,
$\Vhi/\Vhio-1$, as a function of {\sffamily\ssz (a)} $\sg$ for
$\sgc\leq\sg<1$ and {\sffamily\ssz (b)} $\se$ for
$\se(\sgc)\leq\se<\se(1)$ fixing $M=0.1$, $N=1$, $\ld=\kp$ and
$\ns=0.965$. We consider EHI (solid lines) or THI (dashed lines).
Values corresponding to $\sgx$ and $\sgc$ {\sffamily\ssz (a)} or
$\sex$ and $\sec$ {\sffamily\ssz (b)} are also depicted. Some of
the parameters of our models -- derived by employing our numerical
and analytic formulae -- are displayed in the Table .}\label{fig1}
\end{figure}
%%%%%%%%%%%%%%%%%%%%%%%%%%%%%%%%%%%%%%%%%

Employing data from the two representative cases depicted in
\Fref{fig1}, we can analyze further the inflationary mechanism of
our models. Namely, {\ftn\sf (i)} the semiclassical approximation,
used in our analysis, is perfectly valid avoiding possible
corrections from quantum gravity, since
$\Vhi^{1/4}(\sgx)=3.3\cdot10^{-3}$
[$\Vhi^{1/4}(\sgx)=2.9\cdot10^{-3}$] for \ehi\ [\thi] is less than
the ultraviolet cut-off scale, $\mP=1$, of theory -- cf.
\Eref{subP}; {\ftn\sf (ii)} any possible running of the quantities
measured at $\Lambda$ is negligible, since the scale $\Lambda$ in
\Eref{Vrc} is found to be $\Lambda=1.6\cdot10^{-3}$
[$\Lambda=9.5\cdot10^{-4}$] for \ehi\ [\thi], i.e., quite close to
the aforementioned $\Vhi^{1/4}(\sgx)$'s; {\ftn\sf (iii)} the
contribution of $\dV$ to $\Vhi$ is quite suppressed, since
$(\dV/\Vhi)(\sgx)=3.6\cdot10^{-5}$ [$(\dV/\Vhi)(\sgx)=10^{-5}$]
for \ehi\ [\thi]; {\ftn\sf (iv)} the criterion of \Eref{kbou}  is
comfortably fulfilled, since ${\Delta
m_{\pha}^2}/{\Hhi^2}\simeq536.5$ [${\Delta
m_{\pha}^2}/{\Hhi^2}\simeq199.6$] for \ehi\ [\thi] and so no
modification of \sEref{ntot}{a} is necessitated.

A prominent aim of the marriage between E- and T-models and HI is
the diminishment of $\ns$ at the acceptable level of
\sEref{data}{a}. The achievement of this goal is readily
demonstrated in \Fref{fig2} where we display the resulting values
of $\ns$ versus $m$ for \ehi\ -- see \sFref{fig2}{a} -- and \thi
-- see \sFref{fig2}{b} -- with the constraints of \Eref{ntot} and
(\ref{data}{\ftn\sffamily b}) being fulfilled. We set $M=0.1$,
$N=1$ and $\ld=\kp$ (solid line), $\ld=2\kp$ (dashed line) and
$\ld=\kp/2$ (dot-dashed line). The observational allowed region of
\sEref{data}{a} is also limited by thin lines. In both cases $m$
turns out to be of the same order of magnitude (around $10^{-6}$)
but the allowed region for \ehi\ is wider than that in \thi. With
fixed $M$ and $N$, which keep $\kp$ roughly unchanged -- cf.
\Eref{lda} --, we observe that as $m$ increases (without change
order of magnitude) the positive contribution in the approximate
part of \Eref{ns} decreases together with $\ns$, which enters the
observationally favored region of \sEref{data}{a}. For even large
$m$ values, $\sgx$ in \Eref{sgx} approaches unity -- since the
second term of the denominator overshadows the first one --, and
so \Eref{ntot} ceases to be satisfied and the various lines
terminate.

%%%%%%%%%%%%%%%%%%%%%%%%%%%%%%%%%%%%%%%%%%%%%%%%%%%%%%%%%%%%%%%%%%%%%
\begin{figure}[!t]\vspace*{-.10in}
\hspace*{-.23in}
\begin{minipage}{8in}
\epsfig{file=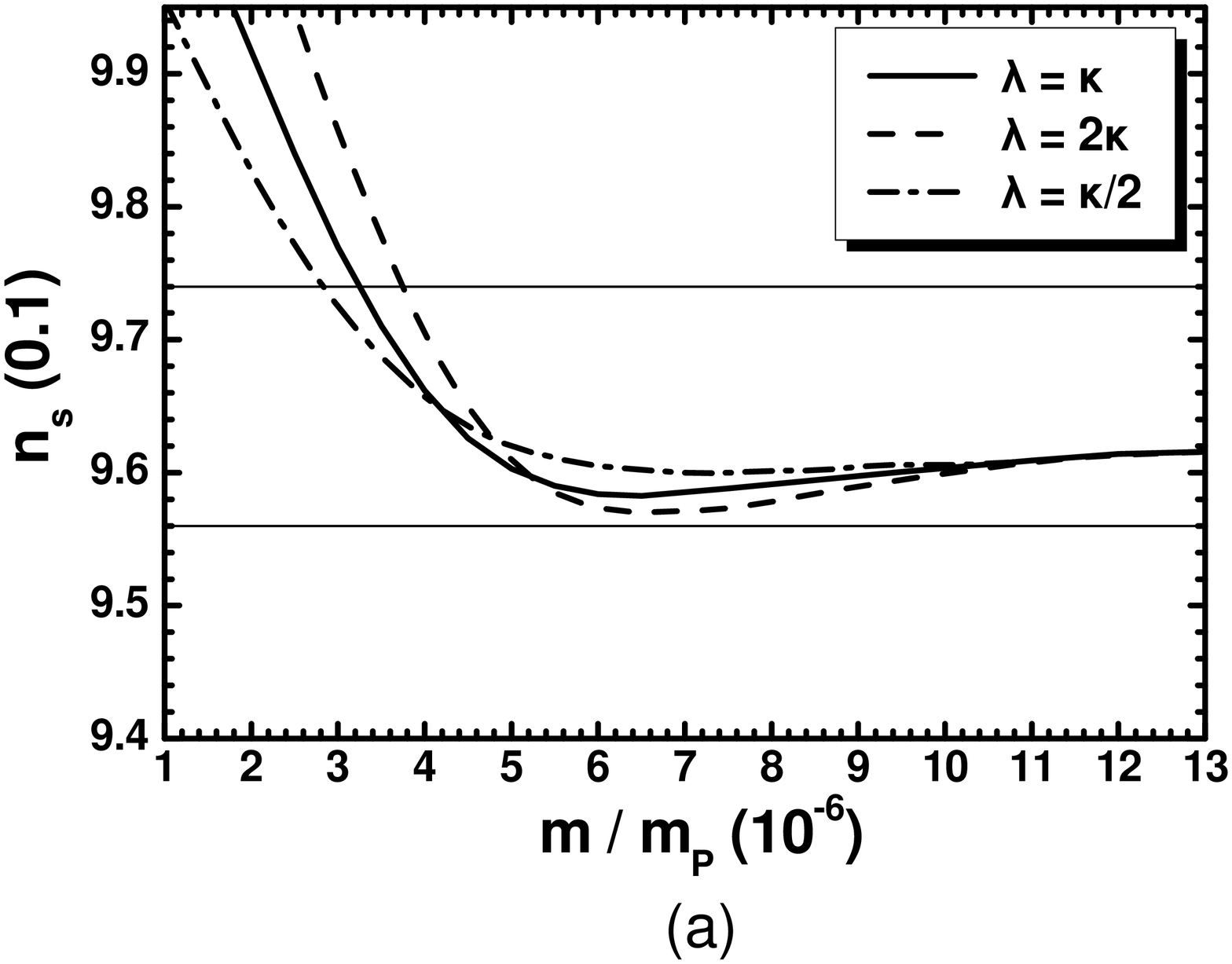,height=3.6in,angle=-90}
\hspace*{-1.2cm}
\epsfig{file=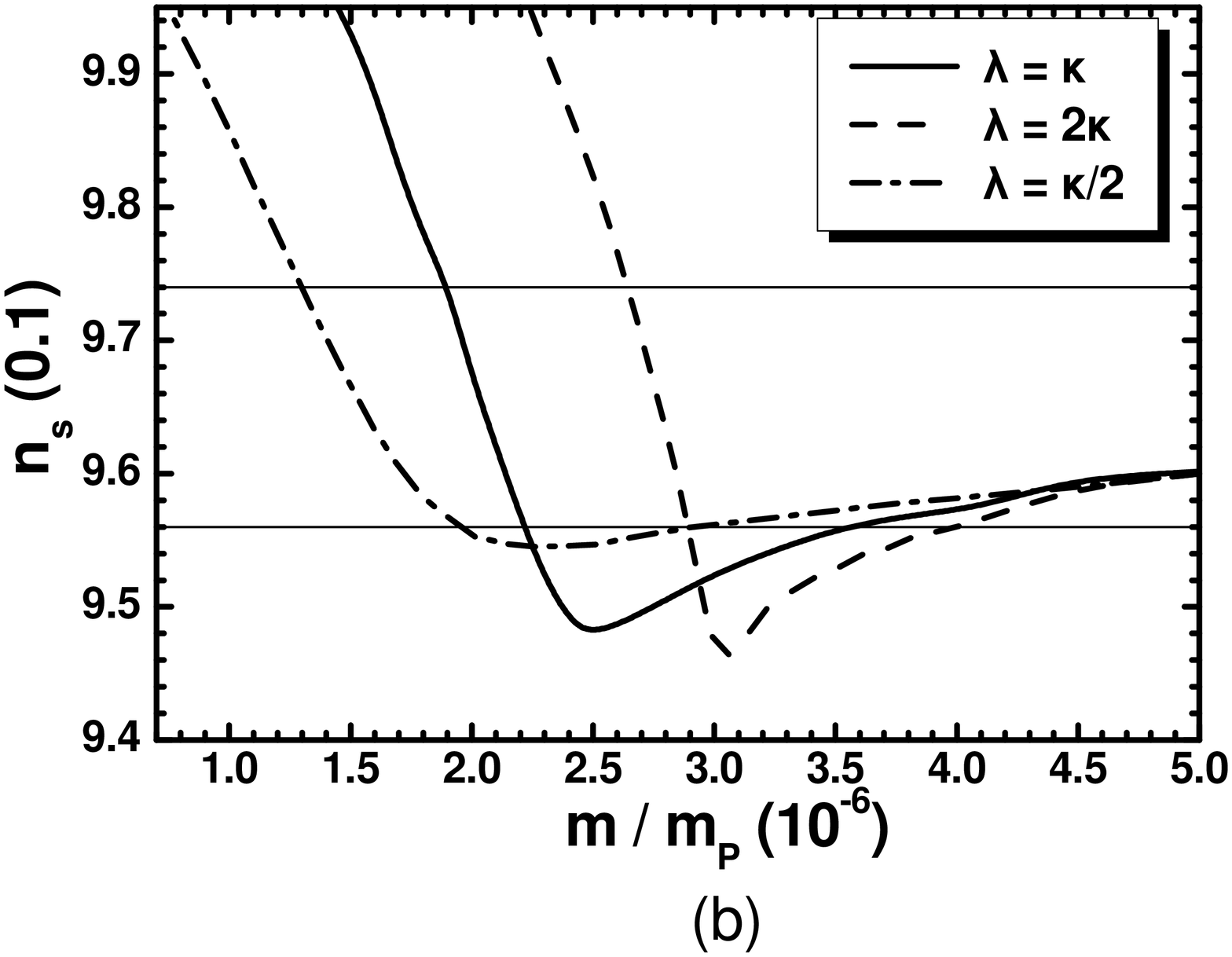,height=3.6in,angle=-90} \hfill
\end{minipage}
\hfill \vchcaption[]{\sl\small The allowed by Eqs.~(\ref{ntot})
and (\ref{data}{\ssz\sffamily b}) values of $\ns$ versus $m$ for
$M=0.1$, $N=1$, several $\lambda/\kp$'s (indicated in the graphs)
and {\ssz\sffamily (a)} EHI or {\ssz\sffamily (b)} THI. The region
of \sEref{data}{\ssz a} is also limited by thin
lines.}\label{fig2}
\end{figure}

%%%%%%%%%%%%%%%%%%%%%%%%%%%%%%%%%%%%%%%%%

%%%%%%%%%%%%%%%%%%%%%%%%%%%%%%%%%%%%%%%%%%%%%%%%%%%%%%%%%%%%%%%%%%%%%
\begin{figure}[!t]\vspace*{-.10in}
\hspace*{-.22in}
\begin{minipage}{8in}
\epsfig{file=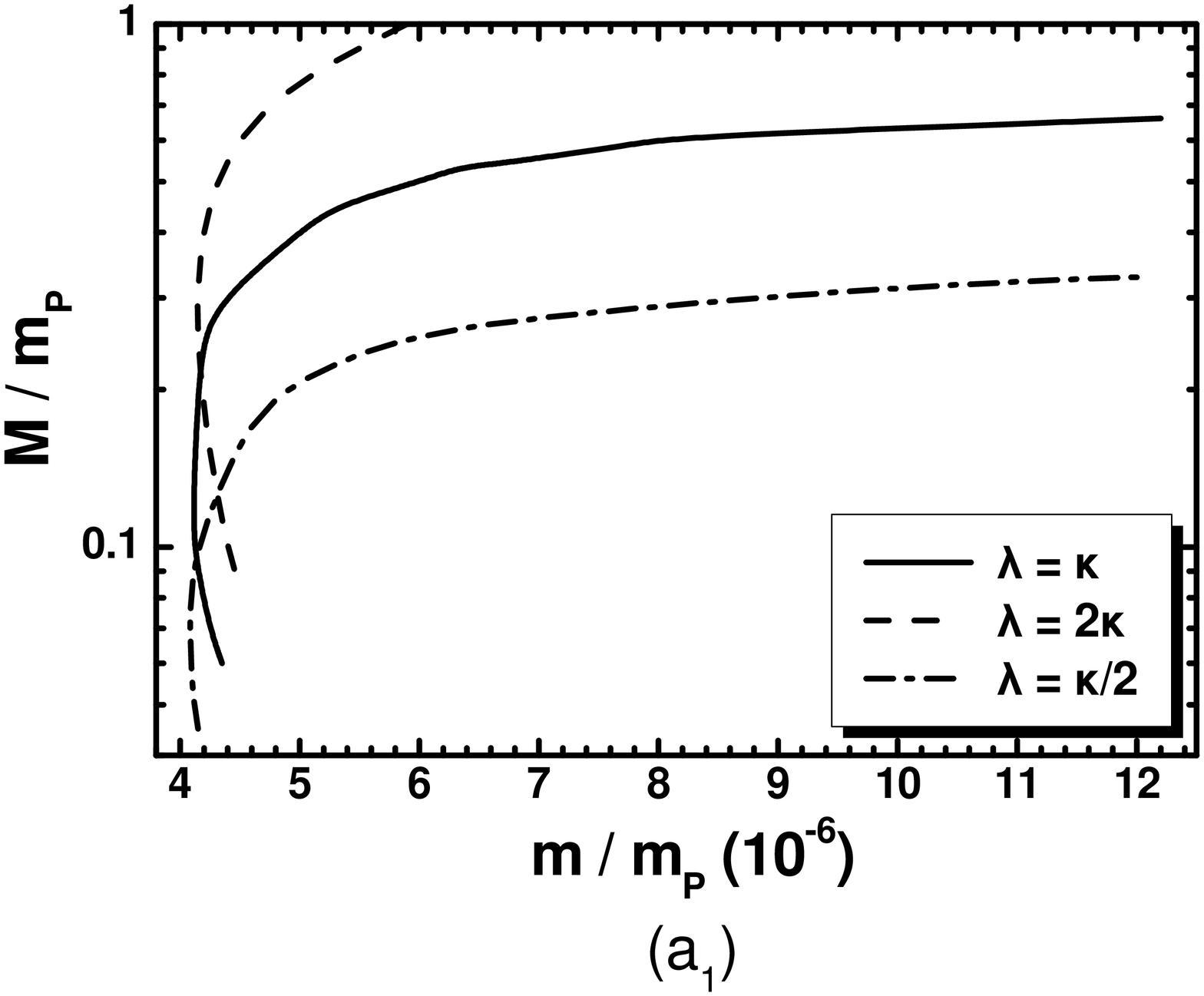,height=3.6in,angle=-90}
\hspace*{-1.2cm}
\epsfig{file=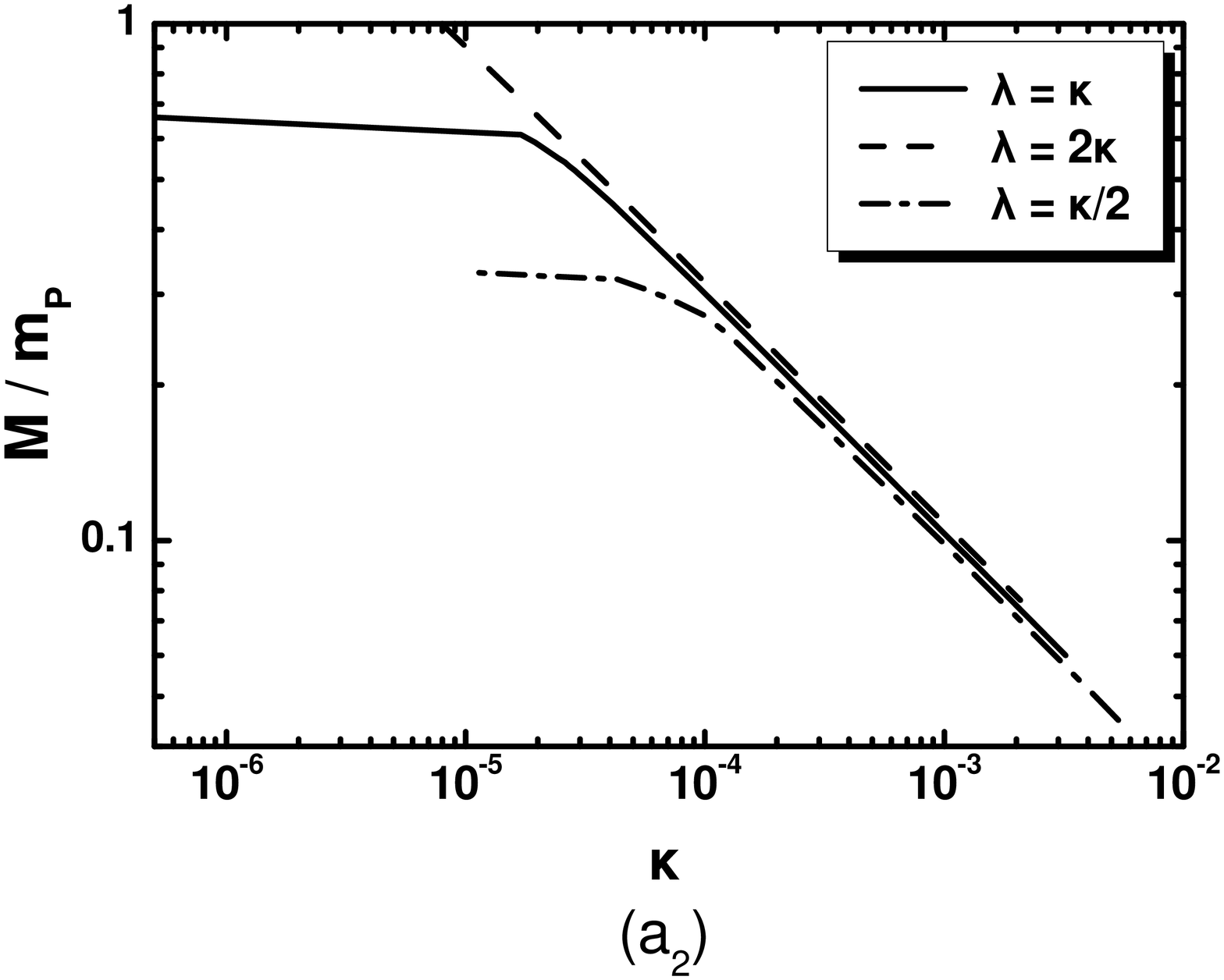,height=3.6in,angle=-90} \hfill
\end{minipage}%\vspace*{-.1in}
\hfill \hspace*{-.22in}
\begin{minipage}{8in}
\epsfig{file=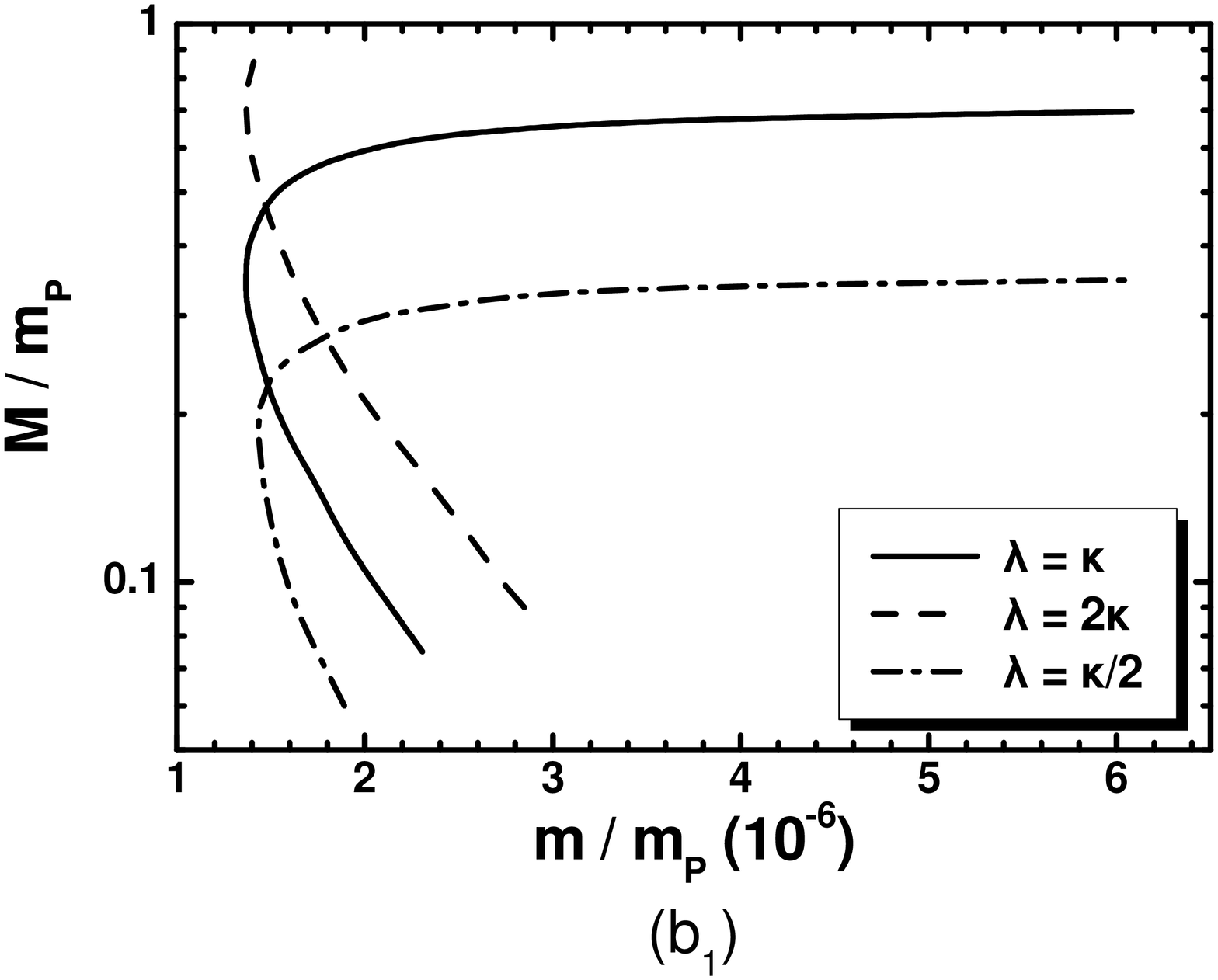,height=3.6in,angle=-90}
\hspace*{-1.2 cm}
\epsfig{file=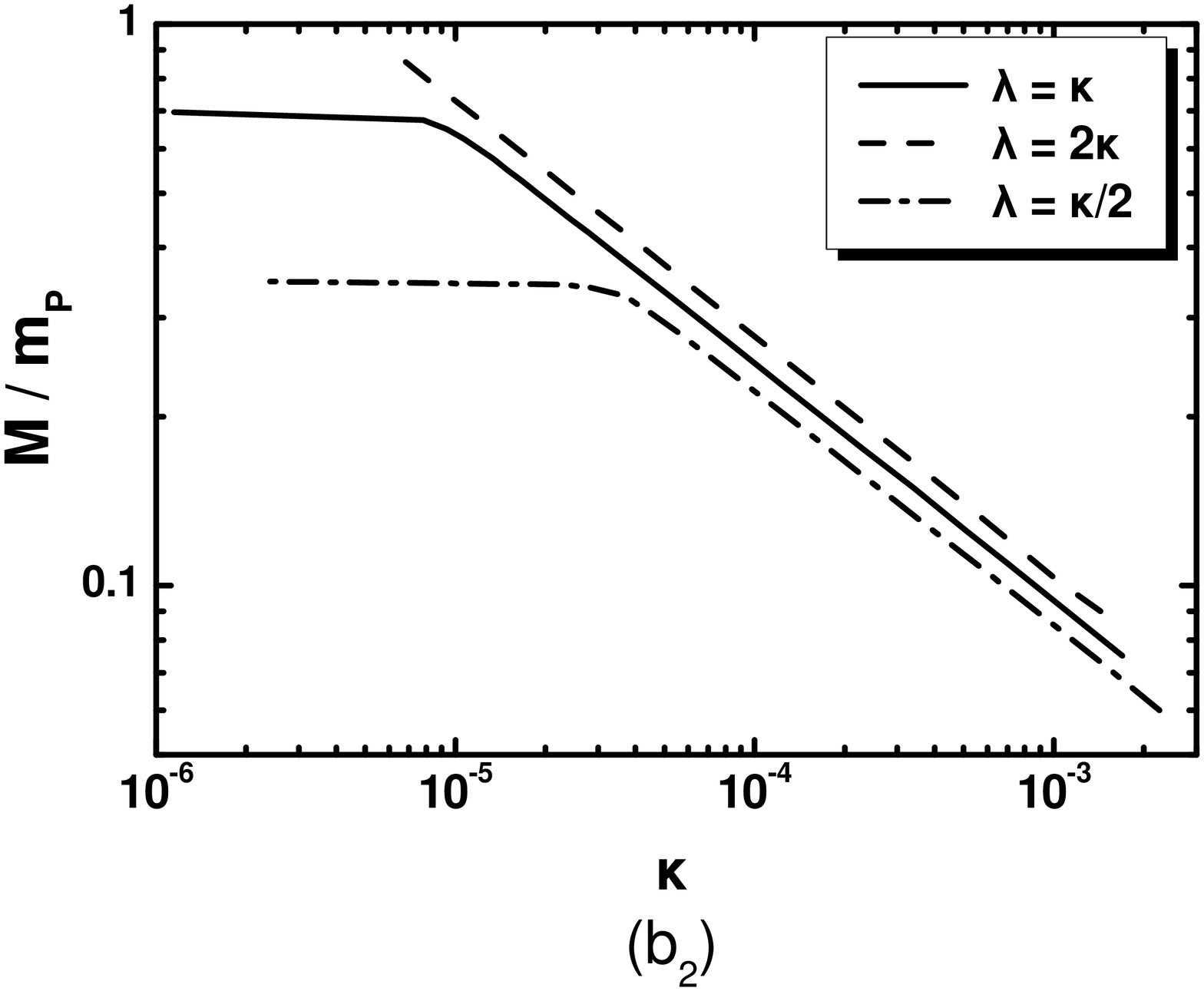,height=3.6in,angle=-90} \hfill
\end{minipage}\hfill\begin{center}\renewcommand{\arraystretch}{1.1}
{\small\begin{tabular}{|c|ccc|cc|c|}\hline {\sc
Model}&$m/10^{12}~\GeV$&$M/10^{17}~\GeV$&$\kappa/10^{-3}$&$\as/10^{-4}$&$r/10^{-3}$&$\Dex~(\%)$\\\hline\hline
\ehi&$9.98 - 29.7$&$1.5 - 16.1$&$0.00047-3.2$&$(-6)-(-2)$&$2.3-4.5$&$1-32$\\
\thi&$3.3 - 14.8$&$1.8 - 17.5$&$0.0011 - 1.7$&$(-6)-10$&$0.6-2.96$&$0.2- 26$\\
\hline
\end{tabular}}
\end{center}%\vspace*{-.08in}
\hfill \vchcaption[]{\sl\small Allowed curves in the $m-M$
[$\kappa-M$] plane ({\ssz\sffamily a$_1$} and {\sf\ssz b$_1$})
[({\ssz\sffamily a$_2$} and {\ssz\sffamily b$_2$})] for $N=1$,
$\ns=0.965$ and various $\ld/\kp$'s indicated in the graphs. We
consider \ehi\ ({\ssz\sffamily a$_1$} and {\sf\ssz a$_2$}) or
\thi\ ({\ssz\sffamily b$_1$} and {\ssz\sffamily b$_2$}). The
allowed ranges of the various parameters for $\ld=\kp$ are listed
in the table with units reinstalled.}\label{fig3}
\end{figure}\renewcommand{\arraystretch}{1.}

%%%%%%%%%%%%%%%%%%%%%%%%%%%%%%%%%%%%%%%%%

Encouraged by the acceptable $\ns$ results above, we proceed to
the delineation of the allowed parameter space of our models
setting $N=1$ and $\ns$ equal to its central value in
\sEref{data}{a} -- possible variation of $\ns$ within its allowed
margin yields relatively narrow strips which would be
indistinguishable especially in the $\kp-M$ plane. We present our
findings in \Fref{fig3} devoting \Fref{fig3}-{\ftn\sf (a$_1$)} and
{\ftn\sf (a$_2$)} to \ehi\ and \Fref{fig3}-{\ftn\sf (b$_1$)} and
{\ftn\sf (b$_2$)} to \thi. In particular, we display in
\Fref{fig3}-{\ftn\sf (a$_1$)} and {\sf\ftn (b$_1$)} the allowed
contours in the $m-M$ plane  and in \Fref{fig3}-{\ftn\sf (a$_2$)}
and {\ftn\sf (b$_2$)} the allowed curves in the $\kappa-M$ plane.
The conventions adopted for the various lines are also shown in
the right-hand side of each graph. In particular, the solid,
dashed and dot-dashed lines correspond to $\ld=\kp$, $\ld=2\kp$
and $\ld=\kp/2$. The various lines terminate at low $M$ (or large
$\kp$) values due to the augmentation of the contribution of $\dV$
in \Eref{Vrc} to $\Vhi$ in \Eref{Vhi}. Indeed, when $|\dV/\Vhi|$
approaches $8\cdot10^{-4}$, $\dV$ starts influence the predictions
of HI. Such a situation is encountered for
$\kp\gtrsim6\cdot10^{-3}$ [$\kp\gtrsim3\cdot10^{-3}$] in \ehi\
[\thi]. On the other side, the almost horizontal part of the
various lines stop for large $m$ (low $\kp$) values since $\sgx$
reaches the pole region -- see \Eref{sgx} -- and the computation
related to \Eref{ntot} becomes unstable. At these points we
encounter the lowest possible $\Dex$'s which tend to the ugly
amount of $0.2\%$. For $\ld=2\kp$ the upper bounds of the
corresponding lines are obtained for $M=1$. We adopt this
conservative bound which assures meaningful $\vev{\pha}$ values
and comfortable fulfillment of \Eref{kbou}.

In the Table of \Fref{fig3}, we accumulate indicatively explicit
values of the various parameters (restoring units hereafter) for
$\ld=\kp$. From the listed values we notice that the allowed $M$'s
approach mainly the string scale contrary to the SUSY versions of
HI where $M$ turns out to be \cite{fhi, su5} close to the SUSY GUT
scale, $M_{\rm GUT}\simeq2.86\cdot10^{16}~{\rm GeV}$. This fact,
though, neither violates \sEref{subP}{a} nor amplifies dangerously
$\dV$ in \Eref{Vhi} since the low enough $\kp$ values, which are
compatible with \Eref{lda}, keep $\Vhi(\sgx)$ and $\dV$ under
control. Comparing the data for EHI and THI, we remark that larger
$m$'s and $\Dex$'s and wider ranges of $\kp$ are allowed in EHI
whereas both models share similar $M$'s. For the same inputs of
the Table \Fref{fig3}, we can estimate the inflaton mass $\msn$
from \Eref{mass}, with results
\beq\label{res1}0.08\lesssim{\msn}/{10^{13}~\GeV}\lesssim
66\>\>\>\mbox{}\mbox{for EHI;}\>\>\> 0.27\lesssim
{\msn}/{10^{13}~\GeV}\lesssim44\>\>\>\mbox{for THI.}\eeq
These ranges let, in principle, open the possibility of
non-thermal leptogenesis \cite{inlept}, if we introduce a suitable
coupling between $\Phi$ and the right-handed neutrinos
\cite{hirc}.

%%%%%%%%%%%%%%%%%%%%%%%%%%%%%%%%%%%%%%%%%%%%%%%%%%%%%%%%%%%%%%%%%%%%%
\begin{figure}[!t]\vspace*{-.1in}
\hspace*{-.23in}
\begin{minipage}{8in}
\epsfig{file=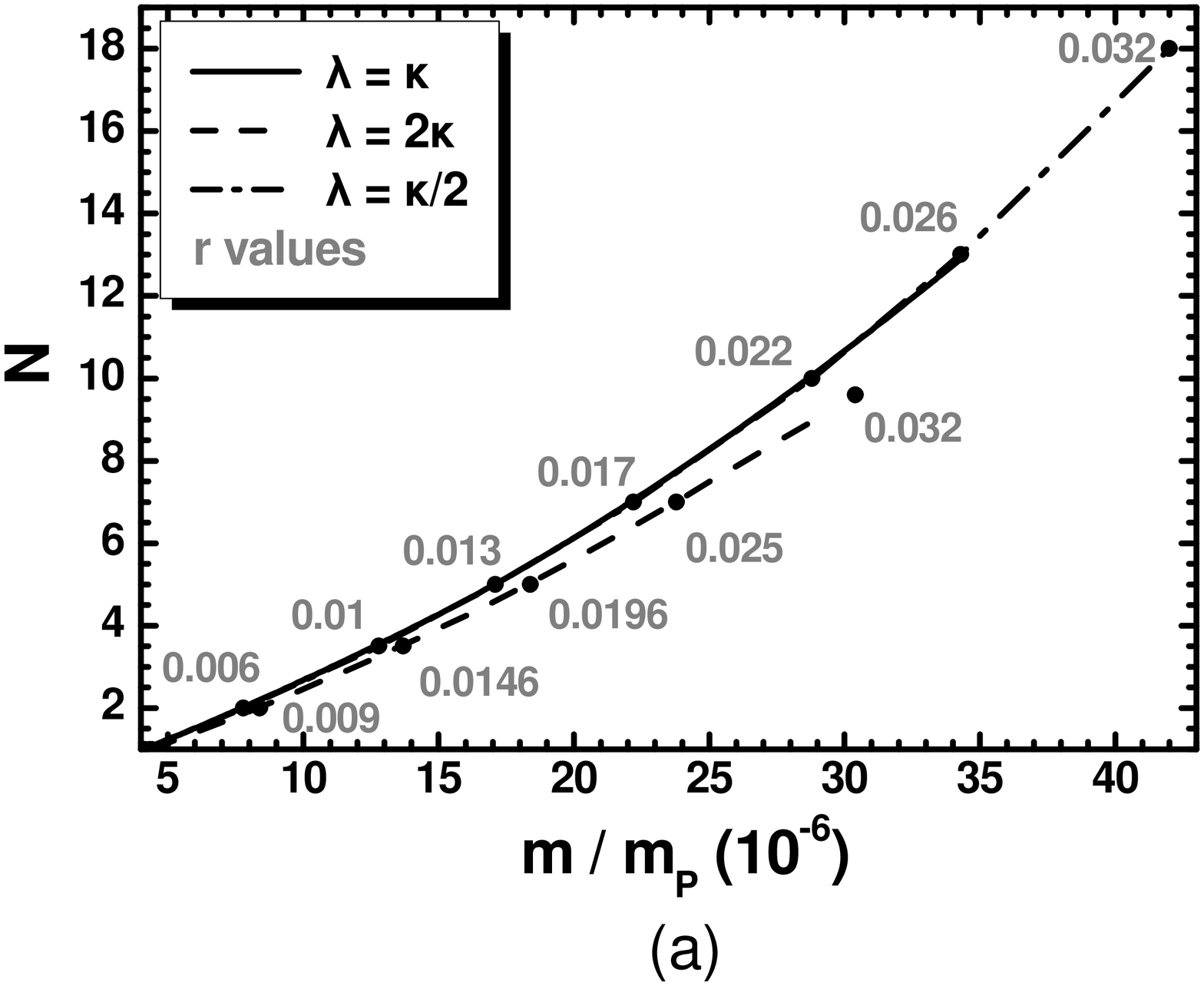,height=3.6in,angle=-90}
\hspace*{-1.2cm}
\epsfig{file=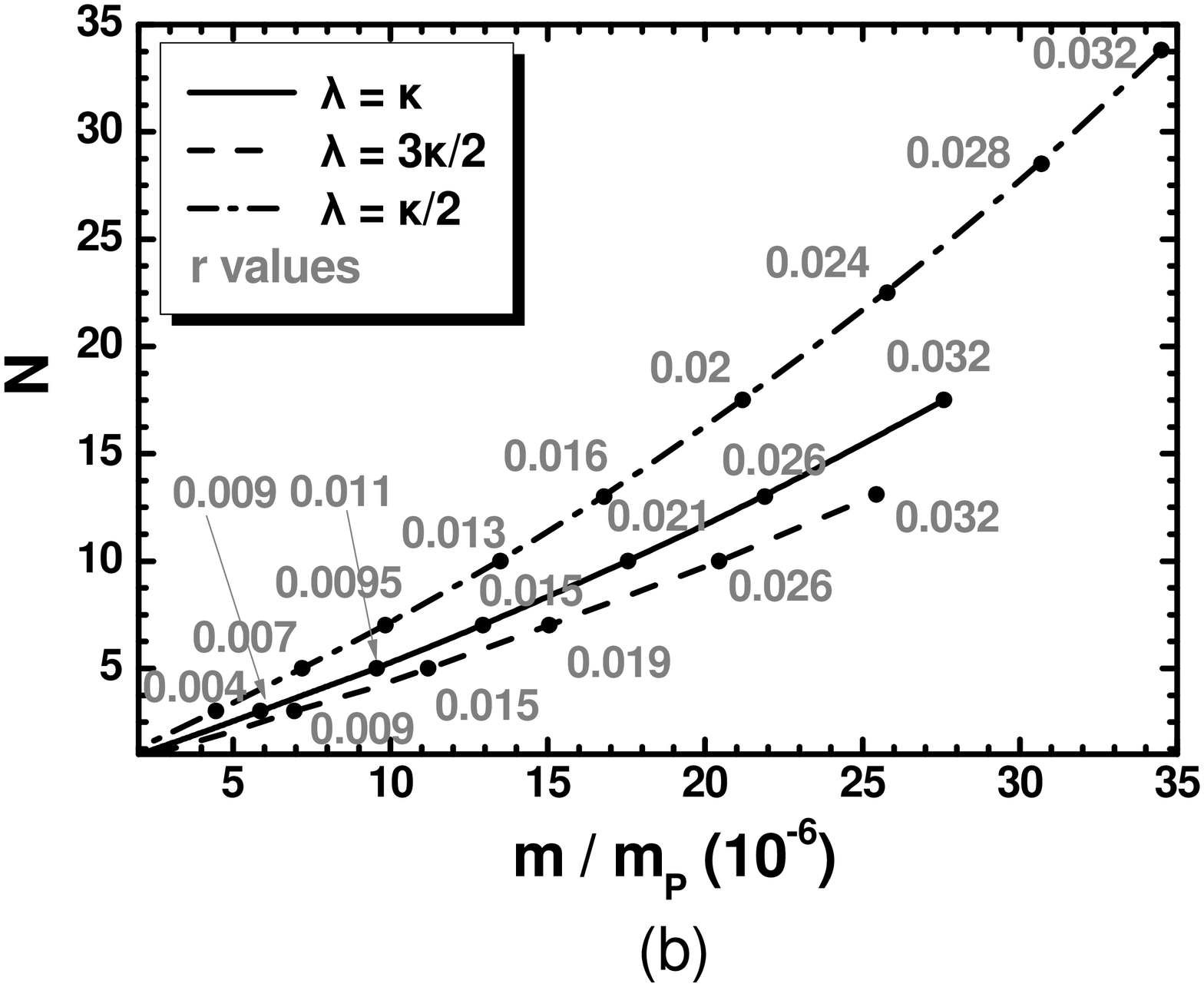,height=3.6in,angle=-90} \hfill
\end{minipage}
\hfill \vchcaption[]{\sl\small  Allowed by \Eref{ntot} values of
$N$ versus $m$ for $M=0.1$, $\ns=0.965$ several $\ld/\kp$'s,
indicated in the graphs, and {\ssz\sffamily (a)} EHI, or
{\ssz\sffamily (b)} THI. Shown is also the variation of $r$ in
grey along the lines. }\label{fig4}
\end{figure}\renewcommand{\arraystretch}{1.}

%%%%%%%%%%%%%%%%%%%%%%%%%%%%%%%%%%%%%%%%%

Varying $N$ beyond unity, we are able to explore another portion
of the parameter space of our models as in \Fref{fig4}. We depict
there the allowed curves in the $m-N$ plane for $M=0.1$, central
$\ns$ in \sEref{data}{a} and \ehi\ -- see \sFref{fig4}{a} -- or
\thi\ -- see \sFref{fig4}{b}. We use the same shape code for the
lines as in \Fref{fig4} with the change that in \sFref{fig4}{b}
the dashed line corresponds to $\ld=3\kp/2$ and not to $\ld=2\kp$
as in all the previous cases. The reason of this replacement is
that RCs invalidate almost the whole parameter space of \thi\ for
$\ld=2\kp$ and $N>1$. From our data we see that $r$ increases with
$N$ as in the standard EMI and TMI \cite{tmodel, etpole} and in
agreement with the analytical findings in \Eref{rs}. The
progressive enhancement of $r$ along each line is shown by gray
numbers. Note that a segment of the solid and dot-dashed lines in
the left graph of \Fref{fig4} coincide between each other. We
display the $r$'s corresponding to $\ld=\kp/2$ along the common
part of these lines. The upper bound on $r$ in \sEref{data}{b}
provides an upper bound on $N$ which is related to the geometry of
the moduli space via \Eref{Rs}. Namely, we find the following
maximal $N$ values
%\hspace*{-2cm}
\beq \label{resN1} N^{\rm max}=\begin{cases}\bem 9.6,&13,&18\\
13.1,&17.5,&33.8\eem\end{cases}\mbox{for}\>\>\>\ld/\kp=\begin{cases}\bem
2,&1,&0.5\>\>\>&\mbox{(\ehi)}\cr
1.5,&1,&0.5 \>\>\>&\mbox{(\thi)} \eem\end{cases}.\eeq
We observe that the largest $N^{\rm max}$ values are obtained for
$\ld<\kp$. This can be understood by the observation that
$\ln\sgc<0$ and dominates the parenthesis in \Eref{rs}. For
$\ld<\kp$, $|\ln\sgc|$ is lower than its value for $\ld\geq\kp$
and so, larger $N$ values are needed so as $r$ to reach its bound
of \sEref{data}{b}. In all, from the data of \Fref{fig4} we find
the following allowed ranges of parameters
\beq\label{resN2} \bem &1\lesssim {m/
10^{13}~\GeV}\lesssim10.2,&1\lesssim
{\kp}/{10^{-3}}\lesssim2.8&\mbox{and}& 1.4\lesssim
{\Dex/0.1}\lesssim3.6&\mbox{for~~EHI;}\\
&3.8\lesssim {m/ 10^{12}~\GeV}\lesssim84,&6.8\lesssim
{\kp}/{10^{-4}}\lesssim29&\mbox{and}& 1.8\lesssim
{\Dex/0.1}\lesssim3.0&\mbox{for~~THI.}\eem\eeq
In the ranges above, we obtain $-6\lesssim\as/10^{-4}\lesssim-1.5$
or $-2.8\lesssim\as/10^{-4}\lesssim6$ for EHI or THI respectively.
We remark that $\Dex$ increases with $N$ (and $r$) in both cases.

%\newpage

\section{Conclusions}\label{con}

Inspired by the successive data releases on CMB perturbations --
which favor the inflationary paradigm -- and the negative LHC
results regarding SUSY, we focused on the well-known model of
non-SUSY HI (i.e. hybrid inflation) \cite{hybrid} attempting to
reconcile it with data. Our proposal is based on the non-SUSY
potential in \Eref{Vhi1} and the non-minimal kinetic terms of the
inflaton sector with a pole of order $p=1$ or $2$ -- see
\eqs{Saction1}{Ksset}. These terms emerge thanks to the adoption
of the logarithmic \Ka s in \Eref{Ket} which parameterize
hyperbolic moduli manifolds. Therefore, our models although
non-SUSY inherit the K\"ahlerian type of moduli geometry from the
SUSY context. This assumption restricts considerably the allowed
$p$ values as shown in Appendix~A. The resulting models are called
\ehi\ (for $p=1$) and \thi\ (for $p=2$) due to the function which
relates the initial and the canonically normalized inflaton -- see
\Eref{je}.

Selecting the gauge group broken after the end of HI and the
representation of the waterfall field we succeeded to evade the
formation of any topological defects. We also included the minimal
possible one-loop RCs (i.e., radiative corrections) to the
inflationary potential and restricted ourselves to the portion of
the parameter space where the waterfall regime occurs suddenly. We
found ample and natural space of parameters compatible with all
the imposed restrictions. E.g., for $N=1$, $\lambda=\kappa$ and
the observationally central value of $\ns$, we find the allowed
ranges shown in the Table of \Fref{fig3}. Increasing $N$ beyond
unity, for the same $\ns$ and $\ld/\kp$ values, we verify an
enhancement on the $r$ values which reaches its maximal allowed
value for $N=13$ or $N=17.5$ within \ehi\ or \thi, respectively.
It is gratifying that no extensive proximity between the values of
the inflaton field at the horizon crossing of the pivot scale and
the location of the pole is needed -- see the $\Dex$ values in the
aforementioned Table which increase with $N$. Therefore, EHI and
THI can be characterized as more natural than the original E- and
T-model inflation as regards this tuning. On the other hand, EHI
and THI are less predictive due to the more parameters which let
corridors for larger variation of the inflationary observables. It
is also remarkable that the one-loop RCs in \Eref{Vrc} do not
affect the inflationary solutions for low enough values of the
relevant coupling constants $\kp$ and $\ld$ -- see \Fref{fig3}.

Comparing our work with that of \cref{hlinde}, we may remark the
following differences-novelties here: {\sf\ftn (i)} We motivated
EHI and THI as non-linear sigma models; {\sf\ftn (ii)} we overcome
the problem of the topological defect through the consideration of
a specific representation of the waterfall field; {\sf\ftn (iii)}
we explored not only \thi\ but also \ehi; {\sf\ftn (iv)} we
considered flat geometry for the waterfall field; {\sf\ftn (v)} we
included the minimal one-loop RCs to the inflationary potential;
{\sf\ftn (vi)} we delineated the parameter space of the models
avoiding any subcritical production of e-foldings. On the other
hand, we did not embed our models in SUGRA as done in
\cref{hlinde}. The standard mechanism \cite{rube} of embedding is
not so effective for HI since the stabilization of the waterfall
fields at the origin does not reproduce the non-SUSY potential and
a rather different situation may emerge. Moreover, the $R$
symmetry preserved by the superpotential \cite{fhi} restricts
further the possible forms of $K$, if it is imposed also to it.
Modification of the present model in order to generate
gravitational waves compatible with the NANOGrav reported signal
\cite{nano}, via a late production of cosmic strings
\cite{kainano} or via a subcritical stage of fast-roll inflation
\cite{dim} could be another interesting target for future
analysis.

%\newpage

\paragraph*{\small \bf\scshape Acknowledgments:} {\small
I would like to thank K. Dimopoulos for useful discussions.
This research work was supported by the Hellenic Foundation for
Research and Innovation (H.F.R.I.) under the ``First Call for
H.F.R.I. Research Projects to support Faculty members and
Researchers and the procurement of high-cost research equipment
grant'' (Project Number: 2251).}

\setcounter{section}{0} \setcounter{subsubsection}{0}
\setcounter{equation}{0}
\renewcommand{\theequation}{A.\arabic{equation}}
\renewcommand{\thesubsubsection}{A.\arabic{subsubsection}}
\renewcommand{\thesubsection}{A.\arabic{subsection}}
\renewcommand{\thesection}{\alphabet{section}}

\appendix
\section*{Appendix A: Kinetic Poles and K\"{a}hler Potentials}\label{math}

%\newpage
%\appendix
\rhead[\fancyplain{}{ \bf \thepage}]{\fancyplain{}{\sl\small E- \&
T-Model Hybrid Inflation}} \lhead[\fancyplain{}{\sl\small
\hspace*{-0.08cm} Appendix A: Kinetic Poles and K\"{a}hler
Potentials}]{\fancyplain{}{\bf \thepage}} \cfoot{}

It is clear that the consideration of a kinetic pole is crucial
for the successful implementation of our proposal. For this reason
it would be interesting to check if the pole order could be
different than the values, $p=1$ and $2$, considered in our work.
In general grounds, the metric of moduli space $K_{SS^*}$ may be a
totally arbitrary real function and so it may include pole of any
$p$. This freedom, though, can be drastically reduced if we
confine ourselves to metrics originated from some \Ka, $K$.

To highlight this key issue, we generalize the $K$'s considered in
\Eref{Ket} adopting the following
\beq
K=-N_K\ln\lf1-\frac13\lf|S|^{2n}+S^qS^{*l}+S^lS^{*q}\rg\rg.\label{Ks}\eeq
The argument of $\ln$ includes a real function $|S|^{2n}$ and a
complex one, $S^qS^{*l}$ -- together with its complex conjugate
which result to a real contribution. We set unity inside $\ln$ to
assure a  well-behaved expansion for low $S$ values, as required
by the consideration of non-renormalizable operators. We insist on
logarithmic $K$'s since these are usually employed in string
theory and assure fractional metrics with possible poles. The
metric $K_{SS^*}$, generated by the $K$ above, along the
inflationary path in \Eref{vevi} is found to be
\beq
\vevi{K_{SS^*}}=N_K\frac{3n^2\sg^{2n}+(l-q)^2\sg^{2(l+q)}+2\sg^{l+q}\lf3lq+(l-n)(n-q)\sg^{2n}\rg}{\sg^2\lf3-\sg^{2n}-2\sg^{l+q}\rg^2}\,.\label{Kss}\eeq
Varying the exponents $n, q$ and $l$ in the domain $(-3,3)$ with
unit step we obtain the metrics
\beq \vevi{K_{SS^*}}=N_K\cdot\begin{cases}1/f_1^2&\mbox{for}\>\>(q,l,n)=(1,0,0)\> \&\> (0,1,0),\\
1/f_2^2&\mbox{for}\>\>\> (q,l,n)=(0,0,\pm1),~ \pm(1,1,0)\> \&\>
\pm (1,1,1),\end{cases}\label{cases}\eeq
which reproduce the -- already used in \Eref{Kc} -- kinetic terms
\beq N\dot \sg^2/2\fp^2\>\>\>\mbox{with}\>\>
\fp=1-\sg^p\label{kin}\eeq
for $p=1$ and $2$. The former is obtained for $N=N_K/2$ whereas
the latter for $N=2N_K$. Note that $\vevi{K_{SS^*}}$ employed in
\thi\ can be derived not only for $n=1$ as expected -- see e.g.
\Eref{Ket} -- but also for $n=-1$. On the other hand, our scanning
does not reveal any other kinetic terms of the type in \Eref{kin}
for $p>2$. The simplest metrics including $\fp$ with $p=3, 4, 5$
and $6$ in the denominator are
\beq
\vevi{K_{SS^*}}=\frac{9N_K\sg^4}{4f_3^2},\>\frac{4N_K\sg^2}{f_4^2},\>\frac{N_K(24+\sg^5)\sg^3}{4f_5^2}\>\>\mbox{and}\>\>\frac{9N_K\sg^4}{f_6^2}\cdot\eeq
It is rather uncertain if the $\vevi{K_{SS^*}}$'s above yield
kinetic terms which support inflationary solutions. As a
consequence, it is highly nontrivial, if not impossible, to
achieve kinetic terms of the form in \Eref{kin} with $p>2$
starting from a \Ka.

%We show below that generalizing the form of $K$ in \Eref{Ket} we
%are not able to derive a kinetic mixing of the form

\def\jcapn#1#2#3#4{{\emph{J. Cosmol. Astropart. Phys.} }{\bf #1}, no.\ #4, #3 (#2)}
\def\prdn#1#2#3#4{{\sl Phys. Rev. D }{\bf #1}, no. #4, #3 (#2)}

\end{document}